\documentclass[%
 reprint,
amsmath,amssymb,
 aps,
]{revtex4-1}

\usepackage{graphicx}
\usepackage{dcolumn}
\usepackage{bm}
\usepackage{epsfig,graphics,amsmath,amssymb}
\usepackage{epstopdf}
\usepackage{upgreek}
\usepackage{booktabs}
\usepackage{color}
\usepackage{draftwatermark}
\usepackage{pifont}
\usepackage{float}

\begin{document}
\preprint{APS/123-QED}

\title{Multidimensional high-harmonic echo spectroscopy:  resolving coherent electron dynamics in the EUV regime}

\author{Shicheng Jiang$^{1}$, Mahesh Gudem$^{2}$, Markus Kowalewski$^{2}$ and Konstantin Dorfman$^{1,3,4\dagger}$ }

\affiliation{$^1$ State Key Laboratory of Precision Spectroscopy, East China Normal University, Shanghai , China}

\affiliation{$^2$ Department of Physics, Stockholm University,  Albanova University Centre, SE-106 91 Stockholm, Sweden}
\affiliation{$^3$ Centerfor Theoretical Physics and School of Sciences,Hainan University,Haikou 570228,China}
\affiliation{$^4$ Himalayan Institute for Advanced Study, Unit of Gopinath Seva Foundation, MIG 38, Avas Vikas, Rishikesh, Uttarakhand 249201, India}

\email[Email:]{$^{\dagger}$ dorfmank@lps.ecnu.edu.cn }

\date{\today}

\begin{abstract}
We theoretically propose a multidimensional high-harmonic echo spectroscopy technique which utilizes strong optical fields to resolve coherent electron dynamics spanning an energy range of multiple electron Volts. Using our recently developed semi-perturbative approach, we can describe the coherent valence electron dynamics driven by a sequence of phase-matched and well separated short few-cycle strong infrared laser pulses. 
The recombination of tunnel-ionized electrons by each pulse coherently populates the valence states of a molecule, which allows
for a direct observation of its dynamics via the high harmonic echo signal. 
The broad bandwidth of the effective dipole between valence states originated from the strong-field excitation results in nontrivial ultra-delayed partial rephasing echo, which is not observed in standard 2D optical spectroscopic techniques. 
We demonstrate the results of simulations  for the anionic molecular system
and show that the ultrafast valence electron dynamics can be well captured with femtosecond resolution..
\end{abstract}

\maketitle
Although the complete information of electronic and nuclear dynamics is encoded in the traditional one-dimensional (1D) optical spectra, the interpretations are challenging as the systems become more and more complex. Increasing the dimensionality of the spectroscopic technique can provide more clues for the analysis of the structural and dynamic information of the targets. Two-dimensional (2D) spectroscopy, originally developed for nuclear magnetic resonance \cite{2DMNR}, has been well-developed in various spectral ranges from the IR \cite{Fleming}, visible \cite{visible2D} to the UV \cite{UV2D}, and has numerous advantages over the 1D spectroscopy in detecting molecular dynamical processes. A more elaborate technique has been obtained by combining two-dimensional (2D) spectroscopy and echo phenomena. Since the first introduction of echo phenomenon in a physical system by Hahn in 1950  \cite{Hahn}, photon echo techniques have since been successfully used to detect ultrafast dynamics in a broad variety of physical and chemical systems \cite{wujianprx,Engel,jcpecho,excitonecho,review}.
Fractional and multiple rotational echoes were observed in numerous experiments  \cite{ilyaprl,ilyapra,wujianprx,wujian}. 
One of the important advantages of photon echo is that the pure dephasig can be detected while the inhomogeneous broadening can be selectively eliminated. 2D photon echo spectroscopy can further decomposes population and coherence dynamics into diagonal and off-diagonal signals in the 2D signal.

\begin{figure*}
\begin{center}
    \includegraphics[width=6.5in,angle=0]{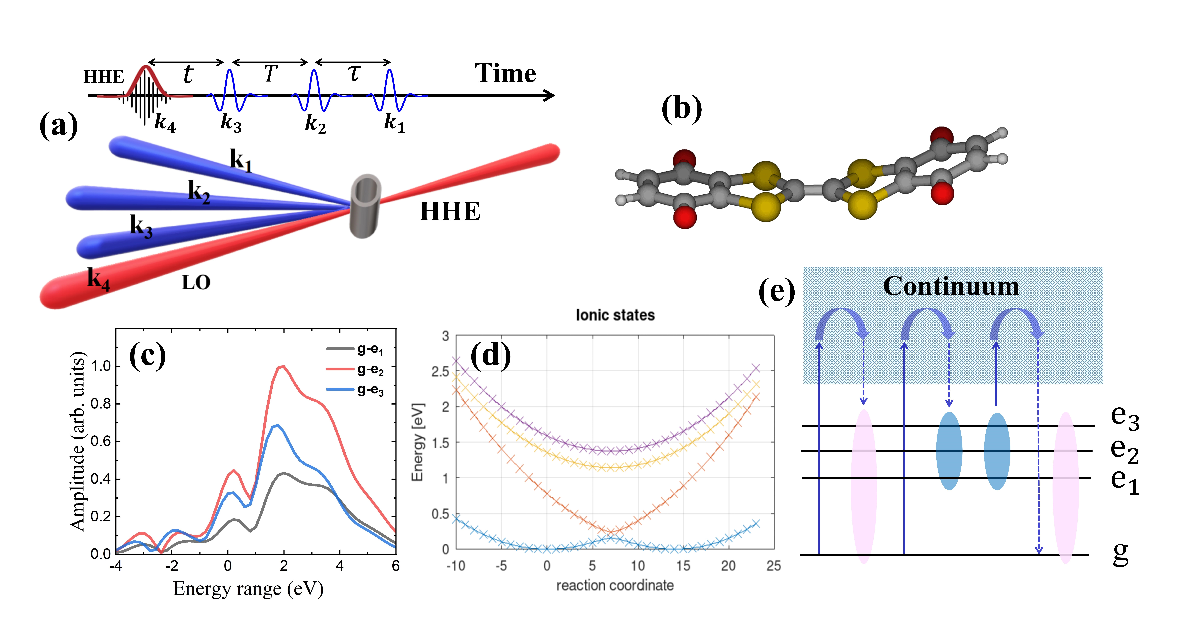}
    \caption{(a) Schematic of the non-collinear multi-pulse setup. The inset panel in the top of (a) is the time order arrangement of the pulses. (b) The optimized structure of Q-TTF-Q$^-$; (c) Fourier transform of the effective transition dipole moment (see Eq. (5) in \cite{jiangOE}) connecting the ground state and the excited states with the maximum value normalized to one; (d) The potential energy curves of Q-TTF-Q$^-$ calculated at CASSCF/6-31G* level of theory. (e) Schematic of the excitation process: the first pulse induces coherences $\rho_{ge}$ between the ground state and excited states through tunneling ionization and the followed recombination; by the same way, the followed second pulse will produce coherences between excited states $\rho_{e_me_n}$ and population $\rho_{e_me_m}$ in each excited state; the third pulse will produce inverse coherences $\rho_{eg}$; finally the delayed echo signal will be emitted automatically.}\label{Fig_1}
\end{center}
\end{figure*}

Existing multidimensional photon techniques and their theoretical description is carried out in the perturbative regime, in which the photon echo signal is generated through a third-order nonlinear process. 
In order to prepare a coherent excitation, the bandwidth of the pump-pulse needs to be sufficiently broad to cover the energy range of the eigenstates.
This becomes important in larger systems, e.g. the biological macromolecules, where
eigenstates are densely distributed and span over a broad range. 

Here, we show that the broadband excitation and the detection of the
subsequent femtosecond dynamics of valence states can be achieved by using strong few-cycle IR pulses.
The presented scheme demonstrates that the use of EUV pulses is not necessary to
cover states in a multi eV range. The use of high harmonics
creates a unique phase matching conditions which allows to suppress the background.
This is not possible with optical or EUV pulses. We further observe an ultra-delayed echo phenomena in the partial rephasing regime..

It has been demonstrated that the strong-field ionization driven few-cycle NIR/IR pulse can produce a coherent superposition of states in the generated cation \cite{5,6,7,8,9,10,11}. 
The opposite process, laser-induced recombination, can in turn 
generate a superposition in the neutral target \cite{16,17}. 
Therefore, strong field techniques can be utilized, in principle, to probe broadband coherences and detect ultrafast dynamics, such as long-range energy transfer and
dephasing of coherences \cite{jiangPNAS}.
We have recently developed a semi-perturbative model (SPM) which treats the strong-field excitation by means of perturbation theory and is capable of describing coherent valence electron dynamics \cite{jiangPNAS, jiangOE}. 
The SPM allows for a separation of processes by different orders, in the same way
time dependent perturbation theory has been used in the optical regime.
Moreover, the SPM allows for a simplification of theoretical description
by using the analytic expansions to describe the interaction between the matter system
and the IR pulses.

Here, we theoretically develop a two-dimensional high harmonic echo (HHE) spectroscopy  in the strong-field regime for the studies of complex molecular systems.The scheme of the spectroscopic setup is shown in Fig. 1(a): three few-cycle laser pulses are incident on the sample along the different directions $\bf{k_1}$, $\bf{k_2}$ and $\bf{k_3}$ to generate HHE. A fourth local oscillator along ${\bf k_4}$ is used for heterodyne detection. In time domain, these fields are defined as $E_j(t-\tau_j)=$exp$({-\frac{(t-\tau_{j})^2}{\sigma^2}})$cos[$\omega_j(t-\tau_j)$] with $\tau_j$ being the center of the $j_{th}$ pulse. The so-called dephasing time $\tau$, waiting time $T$ and rephasing time $t$ are defined as the center to center distances between the pulse pairs: $\tau=\tau_2-\tau_1$, $T=\tau_3-\tau_2$, $t=\tau_4-\tau_3$, respectively. The 2D HHE spectrum can be obtained by 2D Fourier transformation of the detected signal with respect to the coherence time $\tau$ and rephasing time $t$. According to the model outlined in \cite{jiangPNAS}, the amplitude of a bound state $\alpha$ excited by the strong-field three-step process is similar to the single photon excitation, except that the transition dipole moment is replaced by a time-dependent effective transition dipole moment $\mu^c(\tau)$. In this case, the amplitude of state $\alpha$ is $\mathrm{C}_{\alpha} \sim \mathrm{i} \int_{-\infty}^{\mathrm{t}} \mathrm{E}(\tau) \mu_{\alpha, \mathrm{g}}^{\mathrm{c}}(\tau) \mathrm{e}^{\mathrm{i} \omega_{\alpha}(\mathrm{t}-\tau)} \mathrm{d} \tau$. Thus the coherence between the excited state $\alpha$ and another bound excited state $\beta$ is proportional to $\left|\tilde{\rho}_{\mathrm{\alpha\beta}}\right| \sim \left|\mathcal{F}_{\mathrm{\alpha}, \mathrm{g}}^*\left(\omega_{\alpha}\right) \mathcal{F}_{\mathrm{\beta}, \mathrm{g}}\left(\omega_{\beta}\right)\right|$, where $\mathcal{F}(\omega)$ is the Fourier transform of $\mathrm{E}(\mathrm{t}) \mu_{\mathrm{\alpha} / \mathrm{\beta}, \mathrm{g}}^{\mathrm{c}}(\mathrm{t})$. Thus, the coherence between the excited states driven by the strong-field three-step process is determined by the Fourier transform of the effective transition dipole moment, which has both positive and negative frequencies. The proposed spectroscopic technique is demonstrated on tetrathiafulvalene-diquinone anion (Q-TTF-Q$^-$), which is a prototypical mixed-valence system where a particular chemical moiety exists in two different oxidation states.
As shown in Fig. 1(b), Q-TTF-Q consists of two identical quinone rings connected by a hetero cyclic (tetrathiafulvalene) compound, and the additional electron of the corresponding anion can be localized on either one of the quinone groups. 
Consequently, the two mirror symmetric equilibrium geometries are possible for the anion. The complexity of electronic description of Q-TTF-Q$^-$ was demonstrated in several theoretical studies \cite{cdft_qttfq,cdft_diab,th_ch1,th_ch2,th_ch3,th_ch4,th_ch5,th_ch6}. 


By using a semi-perturbative approach, we can classify the response as a combination of rephasing (R), non-rephasing(NR), and ultra-delayed partial rephasing (PR) contributions. 
We can show that both the excited state population decay rate and the coherence dephasing between the excited states can be detected through the two-dimensional HHE. 
The ultra-delayed PR signal allows for obtaining a spectroscopic response originating from a selected fraction of the excited states rather than the entire band of valence states.
It is delayed by an additional time interval compared to the R and NR contributions. 

Following our previous work\cite{jiangPNAS}, the master equation including strong-field three-step excitation reads 
\begin{equation}
\frac{\partial \rho_{\mathrm{mn}}(t)}{\partial t}=-i \sum_k\left[H_{m k}(t) \rho_{k n}(t)-\rho_{m k}(t) H_{n k}^*(t)\right],
\end{equation}
where the effective field–matter interaction Hamiltonian is given by
\begin{equation}
H_{m k}(t)=-E(t)\left(\mu_{m k}^b +\mu_{m k}^c(t) \right) e^{i\left(E_m-E_k\right) t}.
\end{equation}
Here, $\mu^b_{mn}$ is the transition dipole moment between the bound states, $\mu^c_{mn}$ is the effective strong-field transition dipole moments describing the laser-induced electronic transition from the bound state $n$ to state $m$ through the continuum, caused by the laser-induced ionization and recombination. The expression for $\mu^c_{mn}$ can be found in Eq. (5) of Ref. \cite{jiangOE}, which is analogue to the polarization of the strong-field three-step model. The form of the master equation Eq. (1) is the same as the  one dealing with multi-level system\cite{shaulbook} except that an effective transition dipole moments $\mu^c$ is included to consider strong-field three-step excitation. Although the intensity of the driving laser is in the tunneling regime, the effective interaction Hamitonian $E(t)\mu^c(t)$ is very small. Therefore, due to the small populations of the excited bound state excited by such an effective dipole Hamiltonian, the bound states transitions can be treated perturbatively, while the free electron degrees of freedom are kept exact. In this way, the structure of the transition moments resembles the weak perturbative regime, while in reality all the transitions are vector potential-dependent and thus contain all orders in the field-matter interactions. According to the semi-peturbative approach\cite{jiangPNAS,jiangOE}, the non-linear polarization can then be expressed as a third order polarization,

\begin{equation}
\begin{split}
&P^{(3)}(t)=\int d \Gamma \int_{0}^{\infty} d t_{3} \int_{0}^{\infty} d t_{2} \int_{0}^{\infty} d t_{1} S^{(3)} \left(t_{3}, t_{2}, t_{1}\right)\\
&E_{3}\left(t-t_{3}-\tau_3\right) E_{2}\left(t-t_{3}-t_{2}-\tau_2\right) E_{1}\left(t-t_{3}-t_{2}-t_{1}-\tau_1\right)
\end{split}
\end{equation}
where the matter response function is given by
\begin{equation}
\begin{split}
&S^{(3)}\left(\mathrm{t}_{3}, \mathrm{t}_{2}, \mathrm{t}_{1}\right)=i^{3} \theta\left(t_{1}\right) \theta\left(t_{2}\right) \theta\left(t_{3}\right)\\
& \times\sum_{\alpha=1}^{4}\left[R_{\alpha}\left(t, t_{1}, t_{2}, t_{3}\right)-R_{\alpha}^{\prime}\left(t_{1}, t_{1}, t_{2}, t_{3}\right)\right],
\end{split}
\end{equation}
with 
\begin{small}
\begin{equation}
\begin{split}
&R_{1}\left(t, t_{1}, t_{2}, t_{3}\right)= \\
&\!\left\langle\eta_{\Gamma}\!\left(\!t\!-t_{3}\!-\!t_{2}\!, \!t_{1}\!\right)\! \eta_{\Gamma}\!\left(\!t\!-\!t_{3}\!,\! t_{1}\!+\!t_{2}\right)
 \!V_{\Gamma}\!\left(t_{1}\!+\!t_{2}\!+\!t_{3}\!\right)\!\eta_{\Gamma}\!\left(\!t\!-\!t_{3}\!-\!t_{2}\!-\!t_{1},\! 0\right)\right\rangle. 
\end{split}
\end{equation}
\end{small}
The $\theta(\tau)$ is the heaviside step function. The remaining terms, e.g. R$_2$,  R$_3$, and  R$_4$ can be written in a similar way, as shown in Eq. (5.28) in Ref. \cite{shaulbook}. The matrix elements of the effective strong-field Raman scattering (SFRS) polarizability are given by 
\begin{small}
\begin{equation}\label{eq:etaGamma}
\begin{split}
\eta^{m n}_{\Gamma}\left(\mathrm{t}, \mathrm{t}^{\prime}\right)=\left(\mu_{m n}^{b}+\mu_{m n}^{c}(t)\right) \mathcal{G}_{mn,mn}(\Gamma, t^{\prime})\\
V^{m n}_{\Gamma}(t, \Gamma)=\mu_{m n}^{b} \mathcal{G}_{mn,mn}(\Gamma, t^{\prime}). 
\end{split}
\end{equation}
\end{small}
$\mathcal{G}_{mn,mn}$ is the Green's function describing the coherence and population dynamics and $\Gamma$ represents the degree of freedom accounting for a static averaging over the energy distribution due to the environment.
In the present work, fluctuations in the transition frequencies are assumed to obey Gaussian distributions like in Ref. \cite{JCP}. The bandwidth that can be covered by the laser-induced recombination is determined by the width of the Fourier transformation of the effective dipole $\chi(q\omega_0)=\int \mu^{c}(t) e^{i q\omega_0 t} d t$, where $\omega_0$ is the central frequency of the fundamental laser. 
Figure 1(c) shows the absolute values of energy-dependent effective transition dipole 
for a 1.5-cycle pulse with a wavelength of 800 nm and the laser intensity of 1$\times$10$^{13}$ W/cm$^2$.
Here, the effective dipole moment's bandwidth is $\sim$ 4 eV. In principle, the energy range covered by the effective dipole moments can reach a few tens of eV.
However, this would require longer wavelengths, which in turn results in 
lower intensity as in the case of conventional HHG spectrum. 

\begin{figure*}[t]
\begin{center}
    \includegraphics[width=6.5in,angle=0]{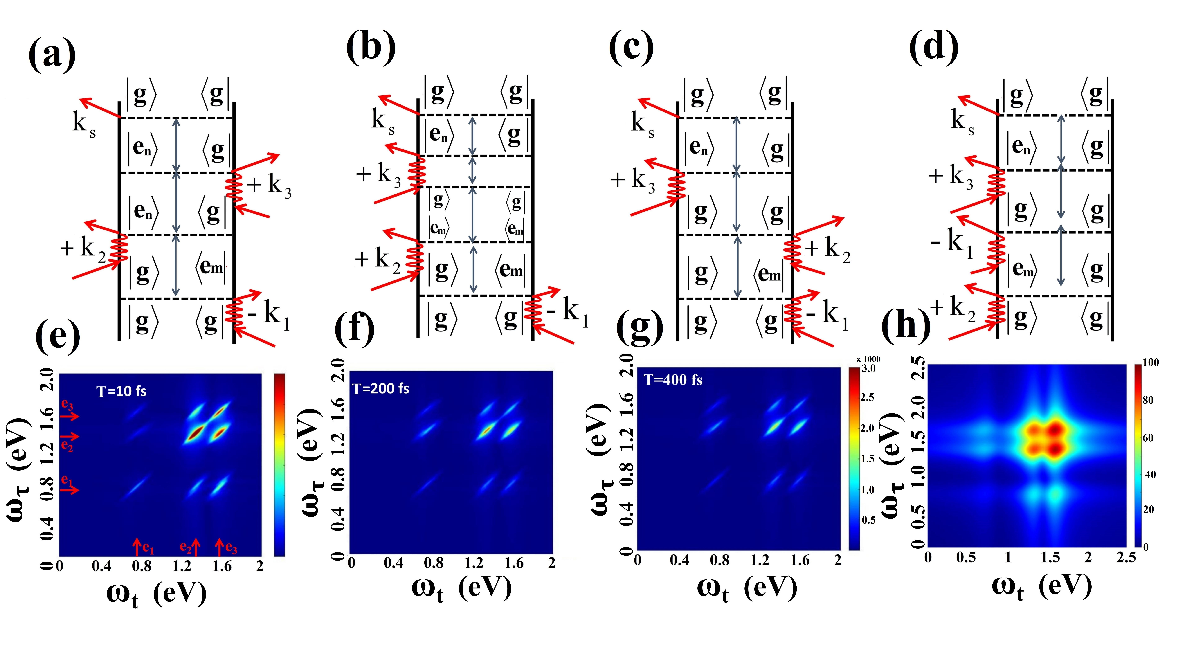}
    \caption{ (a)-(c) The double-sided Feynman diagrams that contribute to the wave-mixing signal $\omega_s=q_1\omega_1+q_2\omega_2+q_3\omega_3$ with $q_1>0$, $q_2<0$ and $q_3<0$.  In each step of field-matter interaction, a wavy line connecting one incoming arrow and one outgoing arrow corresponds to the transition from the bound state to the continuum via the tunneling ionization, and followed by the recombination into another bound state with high harmonic photons emitted. If the incoming arrow is longer, it represents a net absorption process. while the shorter incoming arrow corresponds to the net emission. (e)-(g) are the 2D HHES for different waiting times $T$. (d) and (h) for non-rephasing signal. In (h), the time delay between pulse 1 ($\bf k_1$) and pulse 3 ($\bf k_3$) is fixed at 10 fs. The rules of the diagrams can be found in Ref. \cite{Biggs}. The diagrams shown here are slightly different from the commonly used ones, since the interactions include the ionization process implicitly. Thus we introduce the wavy lines between the arrows pointing inwards (absorption) or outwards (emission) to represent the free electron propagation in continuum.If the arrows pointing inwards are longer than the arrows pointing outwards, it means the whole process absorbs net photon, and vice versa.
}\label{Fig_2}
\end{center}
\end{figure*}

The Green's function in Eq. \ref{eq:etaGamma} then reads \cite{chemreview}:
\begin{small}
\begin{equation}
\begin{split}
&\mathcal{G}_{m n, k l}(\Gamma,t)=\delta_{m n} \delta_{k l}[\exp (-{\hat K} t)]_{m n, k l} \\
&+\left(1-\delta_{m n}\right) \delta_{m k} \delta_{n l} \exp \left(-i \omega_{m n}(\Gamma) t-\gamma_{m n} t\right),
\end{split}
\end{equation}
\end{small}
where $\hat K_{mm,nn}$, $\omega_{mn}$, and $\gamma_{mn}$ represent the population transfer rate, energy difference, and the pure dephasing rate between the states $n$ and $m$, respectively. The numerical values of these parameters can be found in Section 2 of the Supporting Information (SI). 
The population transfer in Q-TTF-Q$^-$ between the two charge transfer states is
mediated by a conical intersection, and thus the  nonadiabatic transfer rate
needs to be taken into account.
To estimate the timescale of population transfer, we solved the Redfield equation\cite{wofgang}, which yields $K_{ee,gg}$$\sim$ -1/172 fs$^{-1}$ for the transfer from the first excited adiabatic state to the ground state.

For simplification, the wave vector can be further introduced by replacing $\omega_st$ by $\omega_st-{\bf k_s}\cdot {\bf r}$ \cite{strelkov}. Since we focus on the strong-field three-step recombination induced excitation, in the following simulation only the strong-field effective transition dipole moments are kept, with $\mu^b$ neglected. The polarization thus reads: 
\begin{equation}
P(r, t)=\sum_{k_{s}, \omega_{s}} e^{\left(i \omega_{s} t-\mathbf{i} \mathbf{k}_{\mathrm{s}} \cdot \mathbf{r}\right)} P\left(\mathbf{k}_{s}, \omega_{s}, t\right).
\end{equation}

The resulting high harmonic wave-mixing signal is then $\omega_s=q_1\omega_1+q_2\omega_2+q_3\omega_3$, with the wave vector ${\bf k_s}=q_1{\bf k_1}+q_2{\bf k}_2+q_3{\bf k_3}$, where $\omega_i$ and $\bf k_i$ are the central frequency and the wave vector of the input pulses, respectively. 
We use identical  central frequencies for the input pulses, 
but chose different propagating directions. 
One can further selectively identify and detect different wave-mixing signals 
by choosing a particular direction for the detection. Usually, in the far-field, the different wave-mixing signals don't have distinct boundaries between each other\cite{Lupx}. Thus it is more realistic to chose  a range of divergency angles.In the following simulations, we only consider signals with $q_1>0$, $q_2<0$ and $q_3<0$. 
We further apply the ideal-time domain approximation in Eq. (1) for simplification\cite{shaulbook}, since the pulse duration is much shorter than the timescale of the system dynamics under consideration. 
To track the field-matter interactions, we select the Feynman diagrams in Fig. 2 (a)-(d) which include only the optical transitions between the ground and the excited states. We have previously extended the diagrams to also include the transitions between the bound states and the continuum \cite{jiangPNAS}.

The first pulse $E_1(t-t_3-t_2-t_1)$ generates a coherence $|g\rangle\left\langle{e}_{m}\right|$ via  SFRS \cite{jiangOE,jiangPNAS}. The corresponding electronic coherences evolve during the coherence time $\tau$. A subsequent, second pulse, then generates a coherence $|e_n\rangle\left\langle{e}_{m}\right|$ between the excited states (when $m \ne n$ in Fig. 2(a)) or the population (Fig. 2(b), (c), (d), and (a) when $m=n$). The waiting time $T$ provides an observation window for the coherence dephasing $\sim e^{-i\omega_{eg}T-\gamma T}$ and the population relaxation $\sim[\exp (-K t)]_{gg,mm}$ dynamics. The third pulse $E_3(t-t_3)$ generates the conjugate coherence which evolves during the rephasing time. The rephasing  allows for a build up of polarization at a certain delay time $t$. 
By scanning the time of $\tau$, $T$, and rephasing time $t$, we can get a three-dimensional signal. The 2D HHE can be obtained by a 2D Fourier transformation with respect to the coherence time $\tau$ and the rephasing time $t$. The population transfer rate from the excited states to the ground state and the excited state coherence dephasing rate in the time interval $T$ can be monitored by visualizing the $T$-dependent diagonal and the off-diagonal peaks in the 2D HHE, respectively.

The electronic properties of the molecular system have been computed by employing the state-averaged CASSCF method \cite{cas1,cas2,cas3} along with a 6-31G* basis set. The potential energy curves (PECs) connecting the two aforementioned minima of Q-TTF-Q$^-$ exhibit an avoided crossing between the lowest two electronic states and are depicted in Fig. 1(d). The transition dipole moments along the PECs and the Dyson orbitals between anionic and neutral Q-TTF-Q have also been calculated at CASSCF/6-31G* level of theory. More details about the molecular system and computational methods are provided in the section 2 of SI.

\begin{figure}[h]
    \includegraphics[width=3.5in,angle=0]{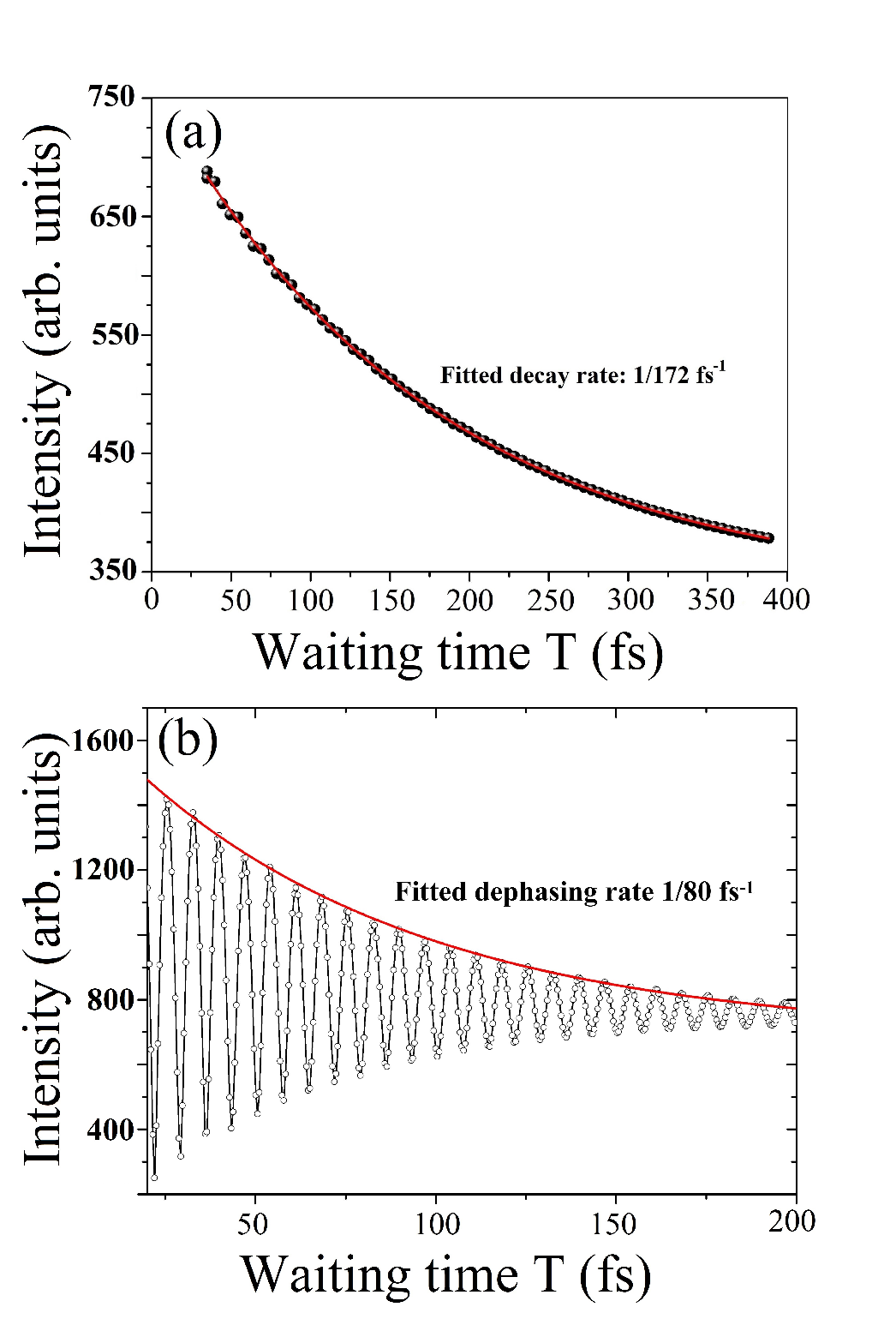}
    \caption{(a), (b): The calculated $T$-dependent signal intensity for diagonal peak $(e_1,e_1)$ and off-diagonal peak $(e_1,e_2)$, respectively. The solid lines are the fitted decay rates.}\label{Fig_3}
\end{figure}

Figures 2 (e) -(f) show the 2D spectrum obtained by a 2D Fourier transformation of $\Re[P(\omega_s,{\bf k}_s,t)]$ along $\tau$ and $t$ for different waiting times $T$.
In most systems, the inhomogeneous broadening is a non-negligible factor contributing to the spectral linewidth.
In the HHE spectra in Fig. 2(e)-(f), the peaks are stretched along the diagonal direction due to a selectively eliminated inhomogeneous broadening along the anti-diagonal direction. In contrast, the NR contribution shown in Fig, 2(h) yields an eroded resolution
due to a large inhomogeneous broadening. 
For the simulation, we have included the electronic ground state and the three electronic
excited states. The molecular geometry has been fixed to the equilibrium geometry
of the electronic ground state.
An extended model including nuclear wave packet motion can be found in our further publication \cite{guanglu}.
Here,  the three excited states result in the corresponding diagonal peaks and the correlation between each pair of excited states leads to the off-diagonal peaks. This behavior is identical
to the traditional photon echo experiments.
When the waiting time $T$ is increased, we observe a decrease of the diagonal peak intensities from Fig. 2(e) to (g).
While, the intensity of the off-diagonal peaks are oscillating with increasing $T$, which can be seen in the Supplementary Movie S1. Figure 3(a) and (b) show the dependence of the intensity of $(e_1,e_1)$ diagonal peak and $(e_1,e_2)$ with respect to the waiting time $T$, respectively.
The fitted decay rate of population transfer $\approx$1/172 fs$^{-1}$ recovers the input parameter calculated via the  Redfield equation. 

So far, we have considered the indirect transition between the ground and the excited states through the continuum. However, transitions between the neighbouring excited states can also occur during the course of interactions with the second and the third pulse. 
The corresponding double-sided diagrams are shown in Fig. 4(a). 
In this case, the time evolution of the coherence during the rephasing process 
is $\sim e^{-i \omega_{e_k e_n} t}$, while it is $\sim e^{-i \omega_{g e_m} \tau}$ during the coherence time. 
The energy differences between the excited states tend to become smaller as the size of
the molecule increases. Moreover, the energy separation between the excited
is often smaller than the energy between the ground and the excited states.
As a result, the build-up of the echo signals will occurs at longer time delays compared to the traditional fully rephased echo signal ($R$ signal), where the revival time of photon echo is equal to the coherence time.
We take the process of the diagram in Fig. 4(a) as an example, by setting $e_m$, $e_n$ and $e_k$ as the first, the second and the third excited state, respectively.  In Fig. 4(b), the coherence time is fixed at 60 fs, while the real part of the polarization builds up at a detecting time $t=200$ fs, which is much longer compared to the $R$ signal. 
Here, it appears well beyond 60 fs and we denote this phenomenon as the “ultra-delayed high harmonic echo" (UDHHE). 
This phenomenon is general and can be understood by the inverse Fourier transformation \cite{swkoch} like $p(t) \propto \int f\left(\omega\right) e^{-i \alpha\omega \tau} e^{i \omega t} d \omega$, where the polarization revival occurs at around $t=\alpha \tau$ ($\alpha\textgreater$ 1, in the above simulations $\alpha\sim 3.3$). 
Fig. 4 (c) illustrates a  2D UDHHE signal, where we
observe a strong cross-peak with partially suppressed inhomogeneous broadening at ($\omega_t,\omega_{\tau}$)=$(e_k-e_n,e_1-e_g)$.

\begin{figure}
    \includegraphics[width=3.5in,angle=0]{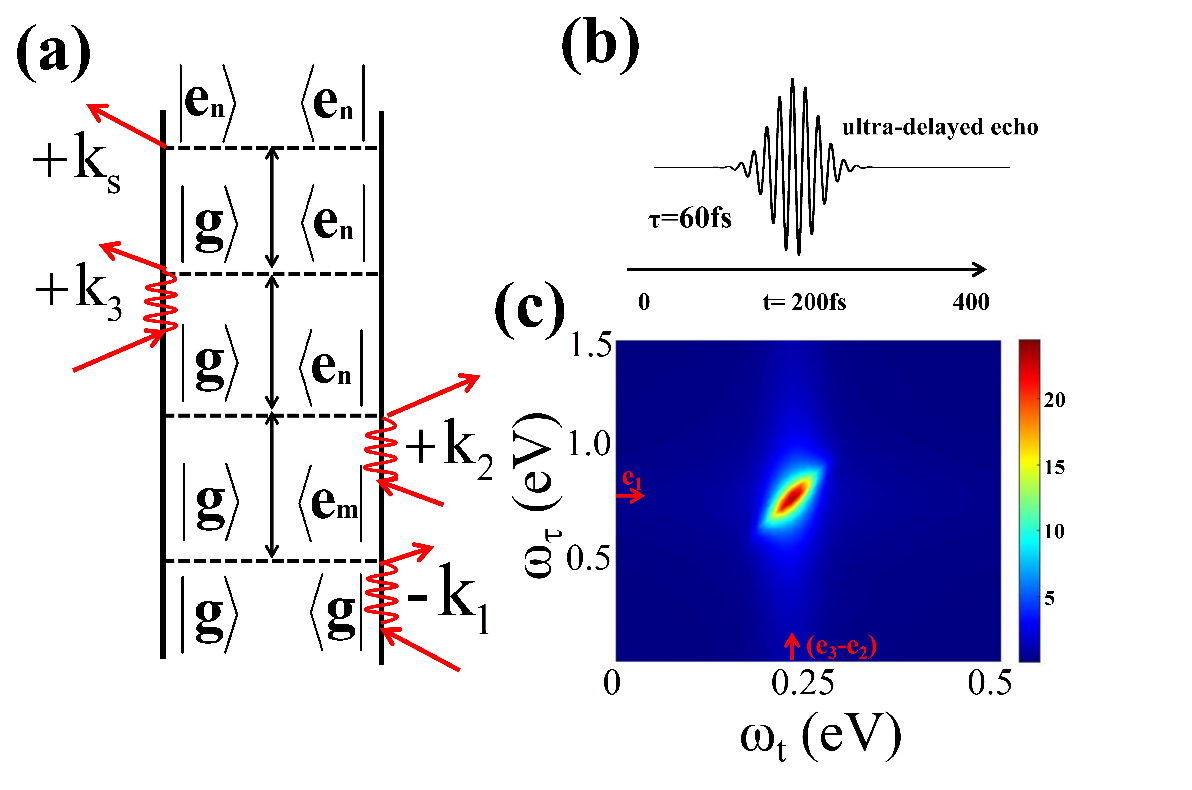}
    \caption{(a) Feynman double-sided diagram for ultra-delayed photon echo signal. (b) The real part of the microscopic polarization depending on the detecting time t. (c) 2D Fourier transformed ultra-delayed echo signal. }\label{Fig_4}
\end{figure}

In summary, we introduced a multidimensional HHE technique that uses strong field
IR pulses rather than EUV pulses to drive transitions between valence electronic states. 
We have demonstrated, that the technique can access a coherent femtosecond dynamics of the electronic states across multiple eV bandwidth without
the requirement of X-ray lasers. We further predict the UDHHE effect, which provides an additional degree of temporal resolution. It can resolve closely spaced electronic states, which otherwise could lead to spectra congestion.
Note, that a vast number of strong field techniques have been used to investigate smaller molecular systems, as they rely heavily on the molecular alignment \cite{alignment}. 
For larger systems, such as biological macromolecules, it is challenging to achieve high degree of alignment, which result in a loss of spectral resolution. 
The proposed HHE technique can eliminate inhomogenous broadening selectively,
and thus may be used for the spectroscopic investigations of biomolecular systems
with increased resolution.

We thank Zhangjie Gao and Junjie Chen for discussion. We acknowledge the support of National Natural Science Foundation of China (12074124); Zijiang Endowed Young Scholar Fund, ECNU; Overseas Expertise Introduction Project (B12024).

\end{document}